\documentclass[11pt,a4paper]{article}

\usepackage{epsfig}
\usepackage{graphicx,epsf}
\usepackage{bm}
\usepackage{amsmath}
\usepackage[english]{babel}
\usepackage{amssymb}
\usepackage{hyperref}

\usepackage{color}

\usepackage[dvipsnames]{xcolor}

\newcommand{\ie}{\emph{i.e., }}
\newcommand{\reff}[1]{(\ref{#1})}
\newcommand{\eref}[1]{Eq.\reff{#1}}
\newcommand{\erefs}[1]{Eqs.\reff{#1}}
\newcommand{\figref}[1]{FIG.\ref{#1}}
\newcommand{\omp}{\omega_p}
\newcommand{\p}{\partial}
\usepackage{color}

\newcommand{\bb}{\scriptscriptstyle{BPS}}
\newcommand{\ee}{\scriptscriptstyle{EGAM}}

\def \d {\mathrm{d}}

\begin{document}

\begin{center}
{\large{\textbf{Nonlinear velocity redistribution caused by energetic-particle-driven geodesic acoustic modes, mapped with the beam-plasma system}}}\\
\vspace{0.2 cm}
{\normalsize
A. Biancalani$^1$, N. Carlevaro$^{2,3}$, A. Bottino$^1$, G. Montani$^{2,4}$, and Z. Qiu$^5$\\
}
\vspace{0.2 cm}
{\footnotesize
$^1$ Max Planck Institute for Plasma Physics, 85748 Garching, Germany\\
$^2$ ENEA, Fusion and Nuclear Safety Department, C. R. Frascati, Via E. Fermi 45, 00044 Frascati (Roma), Italy\\
$^3$ LTCalcoli Srl, Via Bergamo 60, 23807 Merate (LC), Italy\\
$^4$ Physics Department, ``Sapienza'' University of Rome, P.le Aldo Moro 5, 00185 Roma (Italy)\\
$^5$ Institute for Fusion Theory and Simulation and Department of Physics, Zhejiang University, 310027 Hangzhou, People's Republic of China\\
}
\vspace{0.2 cm}
\small{
\vspace{0.2 cm}
contact of main author: \url{www2.ipp.mpg.de/~biancala}}
\end{center}

\begin{abstract}
The nonlinear dynamics of energetic particle (EP) driven geodesic acoustic modes (EGAM) in tokamaks is investigated, and compared with the beam-plasma system (BPS). The EGAM is studied with the global gyrokinetic (GK) particle-in-cell code ORB5, treating the thermal ions and EP (in this case, fast ions) as GK and neglecting the kinetic effects of the electrons. The wave-particle nonlinearity only is considered in the EGAM nonlinear dynamics. The BPS is studied with a 1D code where the thermal plasma is treated as a linear dielectric, and the EP (in this case, fast electrons) with an n-body hamiltonian formulation.
A one-to-one mapping between the EGAM and the BPS is described. The focus is on understanding and predicting the EP redistribution in phase space. We identify here two distint regimes for the mapping: in the low-drive regime, the BPS mapping with the EGAM is found to be complete, and in the high-drive regime, the EGAM dynamics and the BPS dynamics are found to differ. The transition is described with the presence of a non-negligible frequency chirping, which affects the EGAM but not the BPS, above the identified drive threshold. The difference can be resolved by adding an ad-hoc frequency modification to the BPS model. As a main result, the formula for the prediction of the nonlinear width of the velocity redistribution around the resonance velocity is provided.
\end{abstract}

\section{Introduction}

Zonal (i.e. axisymmetric) flows, associated to zonal radial electric fields, are known to exist in tokamak plasmas due to the nonlinear generation by drift-wave turbulence. Both zero-frequency zonal flows (ZFZF)~\cite{Hasegawa79,Rosenbluth98,Diamond05} and finite frequency geodesic acoustic modes (GAM)~\cite{Winsor68,Zonca08,Qiu18} can be excited.
As a consequence of this energy flow from microscopic to mesoscopic scales, ZFZFs and GAMs play a role as major turbulence saturation mechanisms. Moreover, in the presence of energetic particles (EP), EP-driven GAMs (EGAM) can be excited due to inverse Landau damping.
EGAMs have been studied theoretically~\cite{Fu08,Qiu10,Qiu11,WangPRL13,Zarzoso13,Miki15,Zarzoso17,Sasaki17,DiSiena18,Qiu18} and experimentally (see for example Ref.~\cite{Horvath16}). The role of EGAMs as possible mediators between EP and turbulence has also been emphasized~\cite{Zarzoso13}.
One of the main effects of EGAMs in tokamak plasmas is the redistribution of the EP population (crf. Ref.~\cite{Zarzoso18} for the implications on the losses of counter-passing EP). In particular, in phase space, this occurs due to nonlinear inverse Landau damping. As a possible consequence, EGAMs might modify the efficiency of the heating mechanism of neutral beam injectors or ion cyclotron heating.

A kinetic model is necessary for theoretically describing the EGAM. One reason is that the EGAM has a frequency of the order of magnitude of the sound frequency $\omega_s = \sqrt{2} c_s/R_0$, with $c_s = \sqrt{T_e/m_i}$ being the sound velocity (with $T_e$ the electron temperature and $m_i$ the thermal ion mass) and $R_0$ being the major radius, and this is comparable to the transit frequency of thermal ions: therefore, resonances with the thermal ions substantially modify the EGAM frequency.
Another reason is that the damping and excitation mechanisms, i.e. respectively the Landau damping and the inverse Landau damping, are intrinsically wave-particle mechanisms. Moreover, resonances with electrons are found to be important for a proper determination of the damping/growth rates of modes of the family of the GAM, and therefore, when a comparison of the theoretical predictions with experiments is desired, kinetic effects of electrons should also be retained~\cite{Zhang10,Novikau17}.
Due to the fact that numerical simulations in 3D real space and 3D velocity space are numerically too demanding for the present computational capabilities, and include much physics which is not interacting with the EGAM due to separation of scales, it is desirable to reduce the model complexity.
Due to the fact that the EGAM frequency is much lower than the ion gyro-frequency, a reduction is possible from 6D to 5D in phase space, with the gyrokinetic (GK) model. This strongly reduces the computational times. Nevertheless, a comparison with even more simplified reduced models is essential to identify the basic physics of the selected instability, and to push towards modelling techniques which can act in real-time, in parallel to a tokamak discharge. Such are 1D reduced models.

In this paper, we investigate the nonlinear dynamics of EGAMs due to wave-particle nonlinearity. 
A strong analogy between the EGAM and the beam-plasma system (BPS)~\cite{Oneil65,Oneil68} exists~\cite{Qiu11}. Although the BPS is basically a mono-dimensional (1D) problem, and the corresponding unstable wave, i.e. the Langmuir wave, lives in a higher frequency domain, nevertheless both instabilities are driven by a suprathermal species (fast ions for the EGAM, fast electrons for the BPS) via inverse Landau damping.
Moreover, although the EGAM is a 2D problem in an equilibrium toroidal magnetic field, its excitation mechanism, i.e. the inverse Landau damping, acts mainly in one direction, namely the direction parallel to the local equilibrium magnetic field. Therefore, once a proper mapping is made, both instabilities can be investigated in terms of an inverse Landau damping problem in a 1D system.
As a consequence, not only the linear dynamics, but also the nonlinear wave-particle dynamics has strong analogies for the two instabilities. In particular, the bounce frequency of the EP which fall trapped into the perturbed electric field is proportional to the square root of the perturbed electric field~\cite{Qiu11,Levin72}, and the saturated electric field is proportional to the square of the linear growth rate~\cite{BiancalaniJPP17,Levin72}.
As a consequence, the question arises whether also the EP redistribution in phase space can be described with similar models for both instabilities.

The comparison of the nonlinear EP redistribution in velocity for the EGAM and the BPS is the problem faced in this paper. 
The EGAM is studied here with the global GK particle-in-cell code ORB5, which was developed in tokamak geometry for electrostatic turbulence studies~\cite{Jolliet07} and now includes the electro-magnetic multi-species extensions~\cite{Bottino11,Bottino15JPP}.
ORB5 has been verified against analytical theory~\cite{Zarzoso14} and benchmarked against GYSELA~\cite{Biancalani14} and GENE~\cite{DiSiena18} for the linear dyanamics of EGAMs. Moreover, the scaling of the saturated electric field of EGAMs with the linear growth rate, for saturation due to wave-particle nonlinearity, has been studied with ORB5~\cite{BiancalaniJPP17}.
The BPS is studied here with a 1D code treating the thermal plasma as a cold dielectric medium and describing the dynamics of the fast electrons as an N-body problem solved with an Hamiltonian formulation~\cite{Carlevaro15,Carlevaro16,Carlevaro18}.
A mapping of the velocity space for the EGAM system and for the bump-on-tail (BoT) paradigm for the BPS is also formulated, allowing to find a one-to-one correspondence between EP redistribution studied in the two problems.

Two regimes are identified here: the regime where the instabilities are weakly driven shows a very good match between the nonlinear EP redistribution observed in the two problems; on the other hand, above a certain threshold in the drive, a difference is found. The difference occurs due to the nonlinear modification of the mode frequency (i.e. the frequency chirping) which exists for the EGAMs, but is not observed for the BPS for the cases of interest.
This frequency chirping is observed here to shift the resonance velocity of the EGAM, whereas the resonance velocity of the BPS remains constant in time. As a consequence, the EP redistribution for strongly driven EGAM is observed to affect a region of the velocity space which slightly moves in time, creating a qualitatively different picture. It is important to note that the scaling of the saturated electric field with the linear growth rate was found to be quadratic in Ref.~\cite{BiancalaniJPP17}, and it does not change at the threshold of the two regimes identified here. Therefore, we can state that the onset of a non-negligible frequency chirping affects the EP redistribution in velocity space but not the scaling of the saturated levels.
The EP redistribution of the EGAM is shown to be recovered with the BPS in the high-drive regime, by adding an ad-hoc frequency modification to the BPS model. As a main result, the formula for the prediction of the nonlinear width of the velocity redistribution around the resonance velocity is provided, and a good match with GK simulations is found.

The paper is organized as follows. In Sec.~\ref{sec:model}, the gyrokinetic model of ORB5 used here for the study of the EGAM is introduced. In Sec.~\ref{sec:equil}, the equilibrium is defined and the linear dynamics of EGAMs is described. The time evolution of the EGAM is shown in Sec.~\ref{sec:nonlinear}. Sec.~\ref{sec:correspondence-BPS-EGAM} is devoted to a description of the analogy between the EGAM and the beam-plasma systems, and the definition of the mapping. The mapping is applied to the prediction of the EP redistribution in velocity space, which is given in Sec.~\ref{sec:EP-redistribution}, together with the discussion of the regimes of validity.

\section{The gyrokinetic model}
\label{sec:model}

The EGAM problem is investigated here with the global nonlinear GK particle-in-cell code ORB5. ORB5 was written for studying electrostatic turbulence in tokamak plasmas~\cite{Jolliet07}, and extended to treat multiple kinetic species (i.e. thermal ions, electrons, EP, impurities, etc) and electromagnetic perturbations~\cite{Bottino11,Bottino15JPP}.
A collision operator is also implemented in ORB5, for the linearized inclusion of inter-species and like-species collisions.
In this paper, electrostatic collisionless simulations are performed with ORB5. Only the wave-particle nonlinearity is considered, by filtering out all $n\ne 0$ modes and pushing only the EP species along the perturbed trajectories (whereas the bulk ions and electrons follow the unperturbed trajectories).

The model equations of the electrostatic version of ORB5 are the trajectories of the markers, and the gyrokinetic Poisson law for the scalar potential $\phi$. These equations are derived in a Lagrangian formulation~\cite{Tronko16PoP}.
The equations for the marker trajectories for the thermal ions and fast ions (in the electrostatic version of the code) are~\cite{Bottino15JPP}:
\begin{eqnarray}
\dot{\bf R}&=&\frac{1}{m_s}p_\|\frac{\bf{B^*}}{B^*_\parallel} + \frac{c}{q_s B^*_\parallel} {\bf{b}}\times \left[\mu \nabla B + q_s \nabla  \tilde\phi  \right]  \label{eq:traj-1}\\
\dot{p_\|}&=&-\frac{\bf{B^*}}{B^*_\parallel}\cdot\left[\mu \nabla B + q_s
  \nabla  \tilde\phi  \right] \label{eq:traj-2}\\
 \dot{\mu} & = & 0 \label{eq:traj-3}
\end{eqnarray}
The set of coordinates used for the phase space is $({\bf{R}},p_\|,\mu)$, i.e. respectively the gyrocenter position, canonical parallel momentum $p_\| = m_s v_\|$ and magnetic momentum $\mu = m_s v_\perp^2 / (2B)$ (with $m_s$ and $q_s$ being the mass and charge of the species).  $v_\|$ and $v_\perp$ are respectively the parallel and perpendicular component of the particle velocity.
The gyroaverage operator is labeled here by the tilde symbol $\tilde{}$. ${\bf{B}}^*= {\bf{B}} + (c/q_s)  {\bf{\nabla}}\times ({\bf{b}} \, p_\|)$, where ${\bf{B}}$ and ${\bf{b}}$ are the equilibrium magnetic field and magnetic unitary vector.

As we are not interested here in comparing the EGAM damping/growth rates with experimental observations, but only in investigating a specific piece of the nonlinear physics of EGAMs, we neglect kinetic effects of the electrons. This is done by calculating the electron gyrocenter density directly from the value of the scalar potential as~\cite{Bottino15JPP}:
\begin{equation}
n_e({\bf{R}},t) = n_{e0} + \frac{q_e n_{e0}}{T_{e0}} \big( \phi - \bar\phi \big)   \label{eq:adiabatic-electrons}
\end{equation}
where $\bar\phi$ is the flux-surface averaged potential, instead of treating the electrons with markers evolved with Eqs.~\ref{eq:traj-1}, \ref{eq:traj-2}, \ref{eq:traj-3}.

We are interested here in the dynamics of zonal perturbations, and we filter out all non-zonal components. Wave-wave coupling is neglected, by evolving the bulk-ion and electron markers along unperturbed trajectories. This means that, in Eqs.~\ref{eq:traj-1}, \ref{eq:traj-2}, \ref{eq:traj-3} for the bulk ions, the last terms, proportional to the EGAM electric field, are dropped.
The nonlinear wave-particle dynamics is studied by evolving the EP markers along the trajectories which include perturbed terms associated with the EGAM electric field. This means that the EP markers are evolved with  Eqs.~\ref{eq:traj-1}, \ref{eq:traj-2}, \ref{eq:traj-3} where the terms proportional to the EGAM electric field are retained.

Finally, the gyrokinetic Poisson equation is~\cite{Bottino15JPP}:
\begin{equation}
 - \sum_{s\ne e}{\bf{\nabla}} \cdot \frac{n_{0s} m_s c^2}{B^2} \nabla_\perp \phi=  \sum_{s\ne e} \int \d W_s  q_s \, \tilde{\delta f_s}  + q_e n_e({\bf{R}},t)\label{eq:Poisson}
\end{equation}
with $n_{0s} = \int dW_s f_{0s}$. The summation over the species is performed for the bulk ions and for the EP, whereas  the electron contribution is given by $q_e n_e({\bf{R}},t)$.
Here $\delta f_s = f_s - f_{0s}$ is the gyrocenter perturbed distribution function, with $f_s$ and $f_{0s}$ being the total and equilibrium (i.e. independent of time, assumed here to be a Maxwellian) gyrocenter distribution functions.
The integrals are over the phase space volume, with $\d W_s =(2\pi/m_s^2) B_\|^* \d p_\| \d \mu$ being the velocity-space infinitesimal volume element. In this paper, finite-larmor-radius effects are considered, for both thermal and fast ions.

\section{Equilibrium and linear EGAM dynamics}
\label{sec:equil}

We consider here the same tokamak configuration adopted in Ref.~\cite{BiancalaniJPP17}, where the scalings of the EGAM nonlinear saturation levels were studied. 
The tokamak magnetic equilibrium is defined by a major and minor radii of $R_0=1$ m and $a=0.3125$ m, a magnetic field on axis of $B_0=1.9$ T, a flat safety factor radial profile, with $q=2$, and circular flux surfaces (with no Grad-Shafranov shift).
Flat temperature and density profiles are considered at the equilibrium. The bulk plasma temperature is defined by $\rho^*=\rho_s/a$, with $\rho_s = c_s/\Omega_i$, with $c_s = \sqrt{T_e/m_i}$ being the sound speed.
We choose $\rho^*= 1/128=0.0078$ ($\tau_e=T_e/T_i=1$ for all cases described in this paper), corresponding to $2/\rho^*=256$.\\
In the case of a hydrogen plasma, we get a value of the ion cyclotron frequency of $\Omega_i = 1.82 \cdot 10^8$ rad/s and a temperature of $T_{i}=$ 2060 eV. The sound frequency is defined as $\omega_s = 2^{1/2} v_{ti}/R$ (with $v_{ti} = \sqrt{T_i/m_i}$, which for $\tau_e=1$ reads $v_{ti}=c_s$). We obtain $c_{s} = 4.44 \cdot 10^5$ m/s. This corresponds to the following value of the sound frequency:  $\omega_{s} = 6.28 \cdot 10^5$ rad/s. 

The energetic particle distribution function is a double bump-on-tail, with two bumps at $v_\| = \pm v_{bump}$, like in Ref.~\cite{Biancalani14,BiancalaniJPP17}, labelled here as $F_{EP}$. In this paper, $v_{bump}=4 \, v_{ti}$ is chosen.  In order to initialize an EP distribution function which is function of the constants of motion only, we neglect the radial dependence of the magnetic field in  $v_\|(\mu,E,{\bf R}) = \sqrt{2(E-\mu B)/m/v_{ti}} \simeq \tilde{v}_\|(\mu,E)$ 
in the Vlasov equation (details are given in Ref.~\cite{Zarzoso14}).
Neumann and Dirichlet boundary conditions are imposed to the scalar potential, respectively at the inner and outer boundaries, $s=0$ and $s=1$.
%
%


\begin{figure}[t!]
\begin{center}
\includegraphics[width=0.44\textwidth]{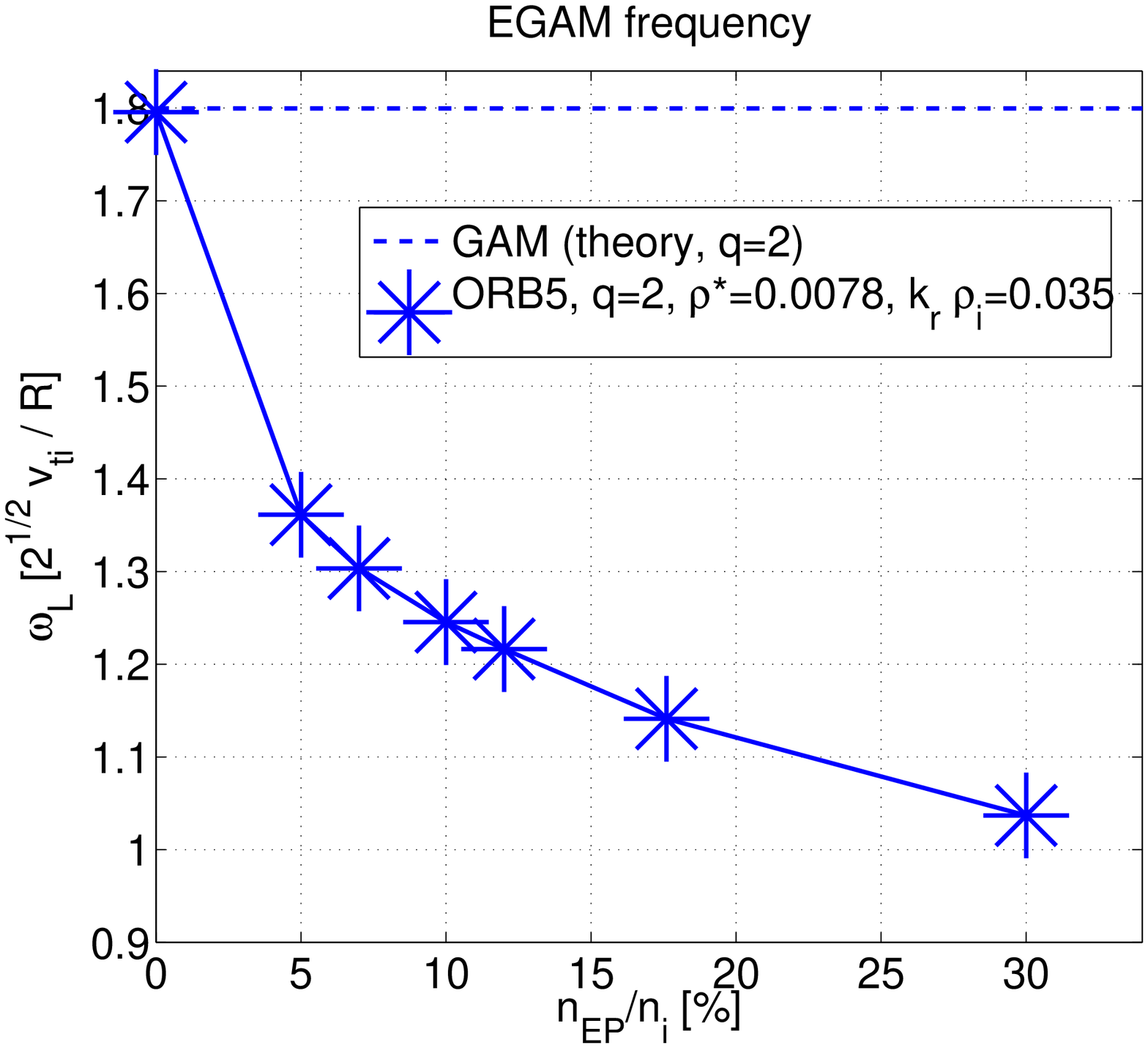}
\includegraphics[width=0.44\textwidth]{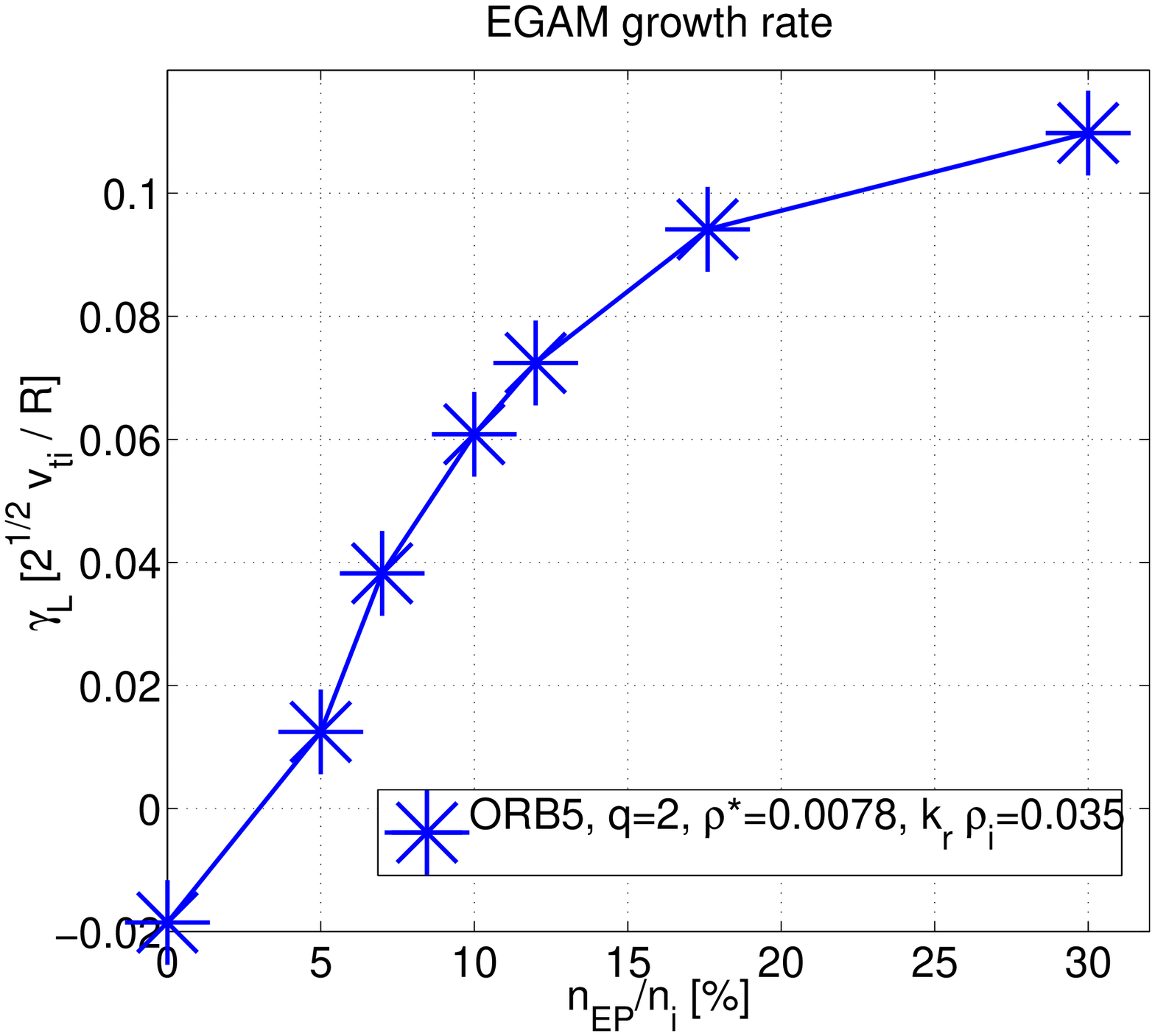}
\caption{Linear frequency (left) and growth rate (right) vs EP concentration.\label{fig:omegagamma_nEP}}
\end{center}
\end{figure}



The scan of linear simulations with different EP concentration, performed in Ref.~\cite{BiancalaniJPP17}, is reported here for completeness. This defines the linear dynamics of the system. The dependence of the linear frequency and growth rate on the EP concentration is shown in Fig.~\ref{fig:omegagamma_nEP}.
For comparison, the GAM frequency for these parameters is $\omega_{GAM}= 1.8 \, \omega_s$.

\section{Nonlinear EGAM evolution}
\label{sec:nonlinear}

\begin{figure}[t!]
\begin{center}
\includegraphics[width=0.51\textwidth]{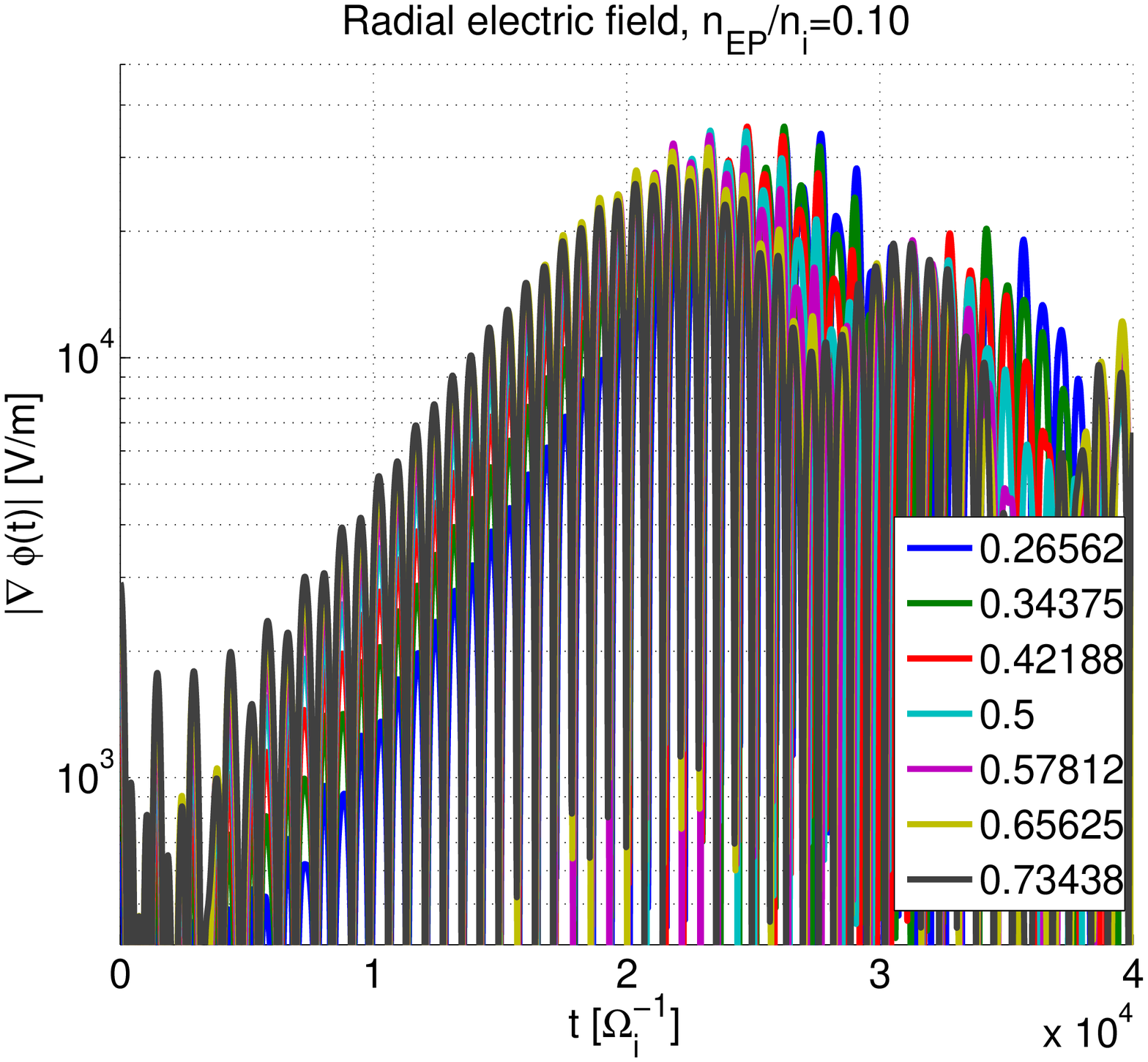}
\includegraphics[width=0.46\textwidth]{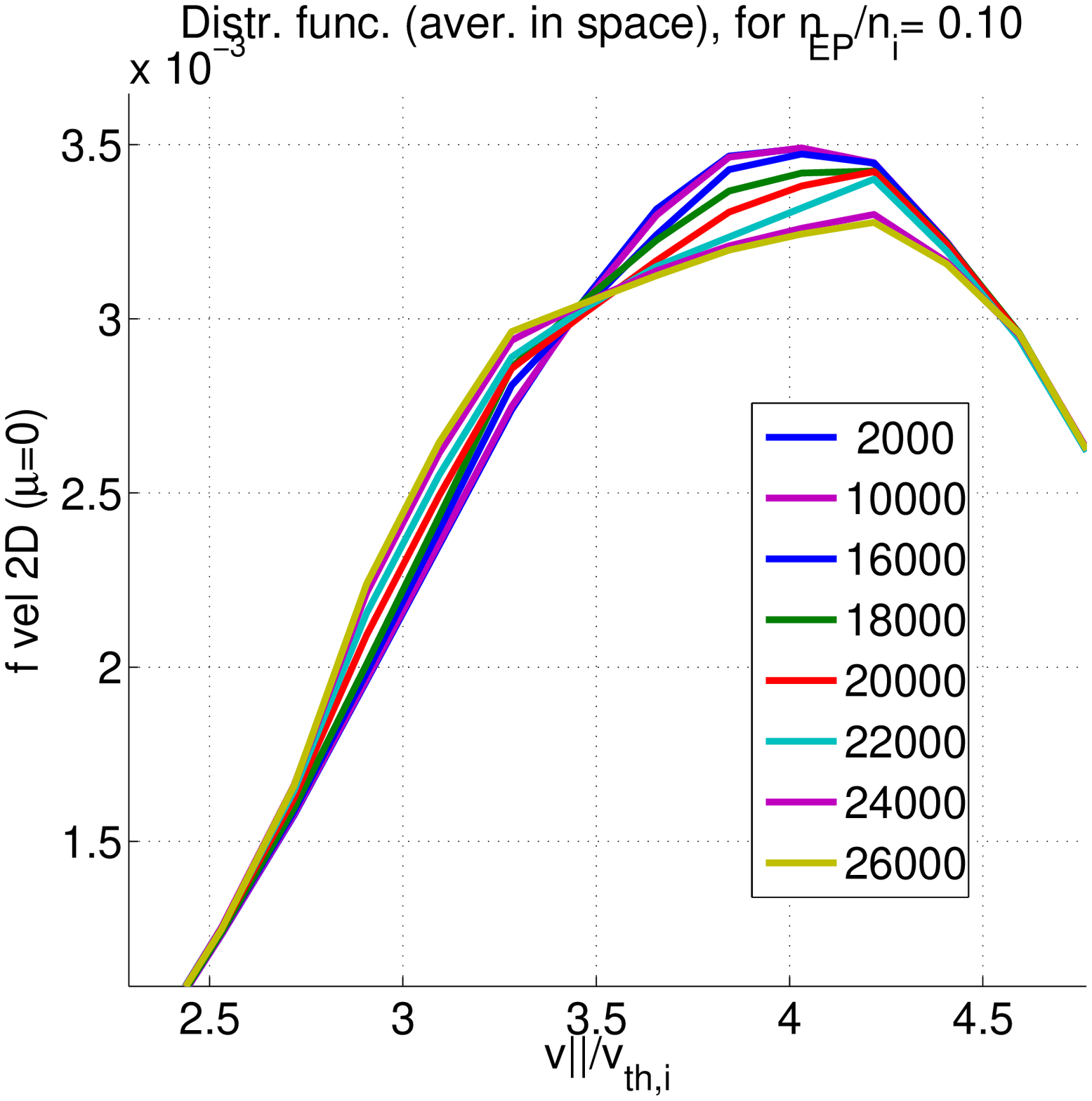}
\vspace{-1em}
\caption{Radial electric field in time, at different radial positions (left), and EP distribution function at different times, vs parallel velocity (right) for an EGAM simulation with ORB5 with  $n_{EP}/n_i=0.10$.\label{fig:nH10-logabsE}}
\end{center}
\end{figure}

In this Section, we describe the evolution in time of the nonlinear simulations of EGAMs performed with ORB5. In this section, like in the rest of this paper, the wave-particle nonlinearity only is considered. As an example, we consider a case with $n_{EP}/n_i=0.10$. A zonal (i.e. axisymmetric) radial electric field is initialized at t=0, with an amplitude of the order of $10^3$ V/m, and let evolve in time in a nonlinear simulation with ORB5. A typical simulation has a spatial resolution set by a grid of (ns,nchi,nphi)=(256, 64, 16) number of points respectively in the radial, poloidal and toroidal direction, a time step of $dt=20 \, \Omega_i^{-1}$, and a number of markers of (ntot$_i$, ntot$_{EP})=(10^7,10^7)$ respectively for the thermal and fast ions.
An initial linear phase is observed, where the radial electric field grows esponentially in time. In this phase the linear frequency and growth rate is measured and checked to match with the ones of the linear simulation:  $\omega_L = 1.24 \, \omega_s$, $\gamma_L = 0.06 \, \omega_s$. Then, a nonlinear phase is entered, the growth rate gradually decreases to zero, and the radial electric field saturates at $t \simeq 2.5\cdot 10^4 \, \Omega_i^{-1}$ (see Fig.~\ref{fig:nH10-logabsE}-a), when the electric field reaches a value of $\delta{E_r} \simeq 3.5 \cdot 10^4$ V/m. This value of the saturated electric field can be compared with the prediction of Ref.~\cite{BiancalaniJPP17}:
\begin{equation}\label{eq:Er-theory}
\delta E_{r,th} = \frac{2 R B \beta_0^2}{\omega_{GAM}} \; \gamma_L^2 = 3.5 \cdot 10^4 \frac{V}{m}
\end{equation}
where the constant $\beta_0 = 2.66$ is estimated in Ref.~\cite{BiancalaniJPP17} for this regime. We emphasize here that the quadratic scaling of the electric field with the linear growth rate shown in Eq.~\ref{eq:Er-theory} has been found to be valid for the whole considered range of EP concentrations (the same range used in Fig.~\ref{fig:omegagamma_nEP}).
After the saturation, the EGAM enters a deep nonlinear phase ($t > 2.5\cdot 10^4 \, \Omega_i^{-1}$), when the electric field starts decreasing in amplitude. In this paper we are interested in the first nonlinear phase only, up to the saturation, and we leave the study of the deep nonlinear phase to another dedicated paper.
In particular, we focus here on the nonlinear modification of the EP distr. funct. at the time of the saturation, and on the corresponding nonlinear modification of the EGAM frequency.

The EP distribution function redistributes in $v_\|$ during the first nonlinear phase, causing a relaxation of the drive due to the inverse Landau damping. The EP distribution function and the EP perturbed distribution function of this simulation is shown in Fig.~\ref{fig:nH10-logabsE}-b). The redistribution of the EPs is observed to occur in a range of velocities between $2.5 \, v_{ti}$ and $4.5 \, v_{ti}$. The EP distribution function does not change during the linear phase, and when entering the nonlinear phase, the redistribution occurs with higher-velocity EP moving towards lower values of $v_\|$, as time increases.
Therefore, negative values of the perturbed distribution function are measured at high velocities, and positive at low velocities. The resonance velocity can be calculated as $v_{\|res} = qR \, \omega_{EGAM} = 3.5 \, v_{ti}$, and can be measured in Fig.~\ref{fig:nH10-logabsE}-b as the velocity where the perturbed distribution function changes sign.
We note that this velocity measured in Fig.~\ref{fig:nH10-logabsE}-b, for this value of $n_{EP}/n_i=0.10$, does not sensibly change in the time range of interest.

\begin{figure}[t!]
\begin{center}
\includegraphics[width=0.49\textwidth]{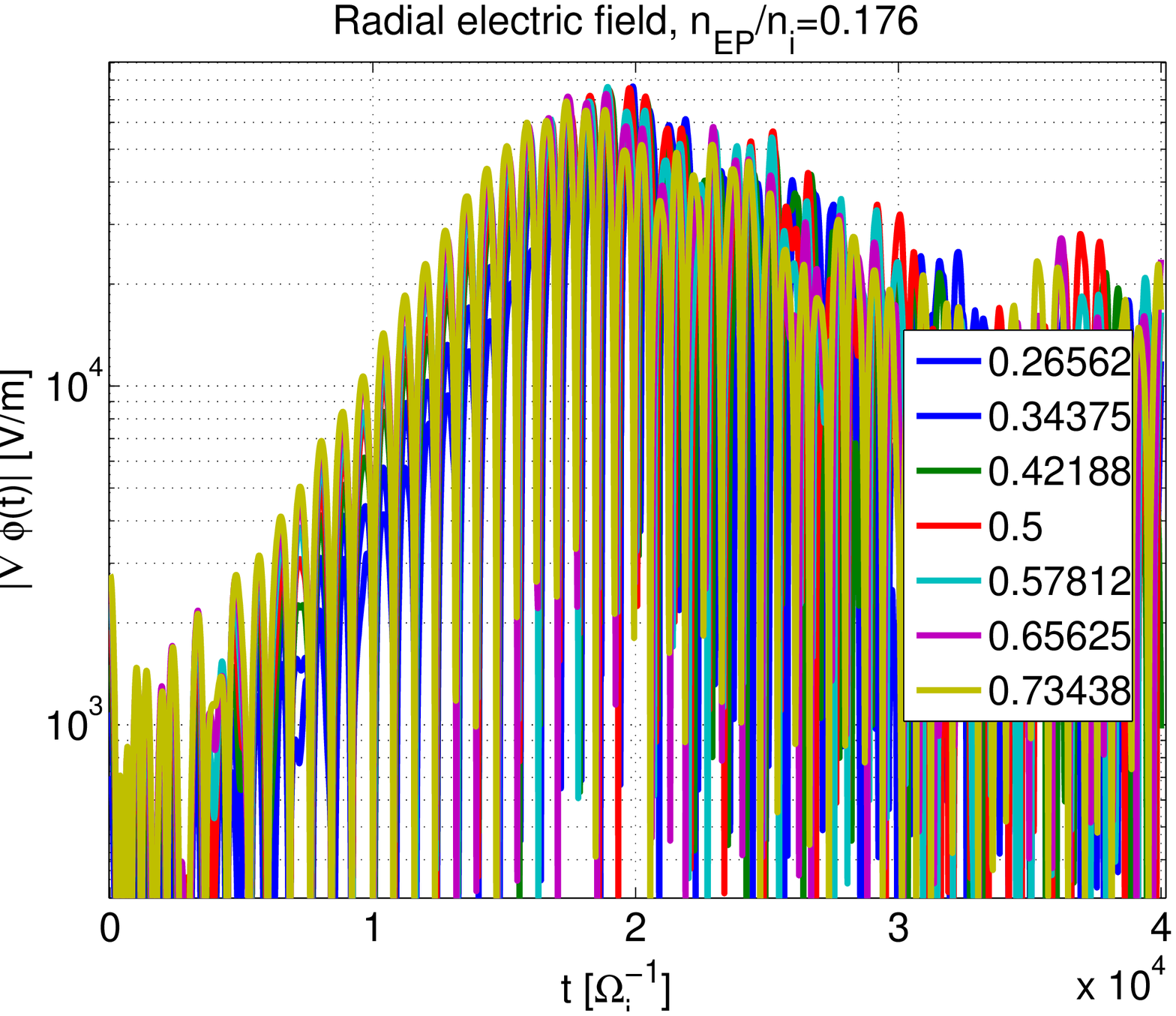}
\includegraphics[width=0.47\textwidth]{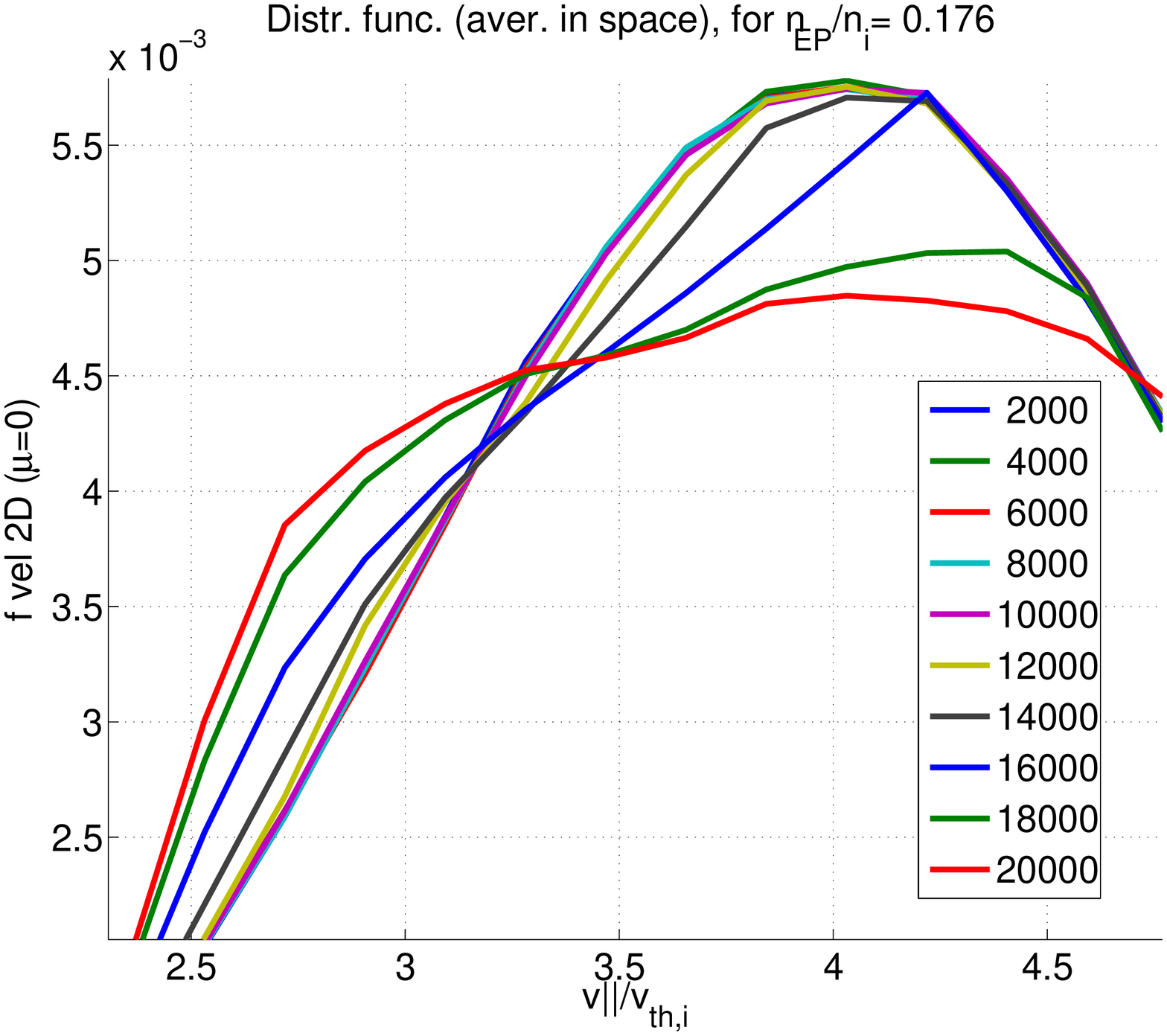}
\vspace{-1em}
\caption{Radial electric field in time, at different radial positions (left), and EP distribution function at different times, vs parallel velocity (right) for an EGAM simulation with ORB5 with   $n_{EP}/n_i=0.176$.\label{fig:nH17-logabsE}}
\end{center}
\end{figure}

Before moving farther, we want to consider another case for comparison, with a stronger drive, namely with $n_{EP}/n_i=0.176$. The evolution in time of the radial electric field and is shown in Fig.~\ref{fig:nH17-logabsE}-a.
The linear frequency and growth rate is measured and checked to match with the ones of the linear simulation:  $\omega_L = 1.14 \, \omega_s$, $\gamma_L = 0.094 \, \omega_s$. The saturated level of the radial electric field is measured at $\delta{E_r} \simeq 0.8 \cdot 10^5$ V/m (in agreement with the prediction of Ref.~\cite{BiancalaniJPP17}).

\begin{figure}[b!]
\begin{center}
\includegraphics[width=0.48\textwidth]{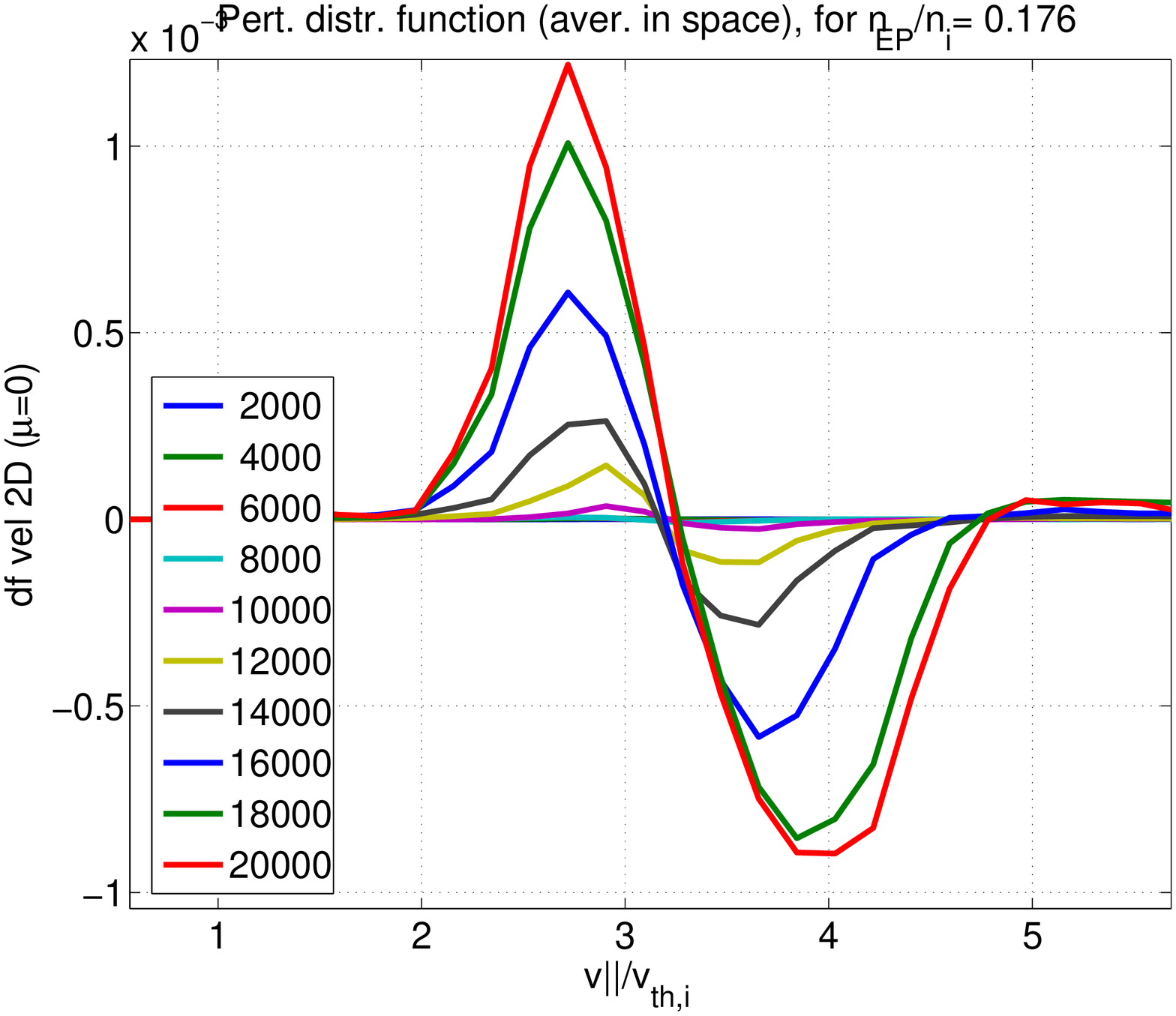}
\includegraphics[width=0.48\textwidth]{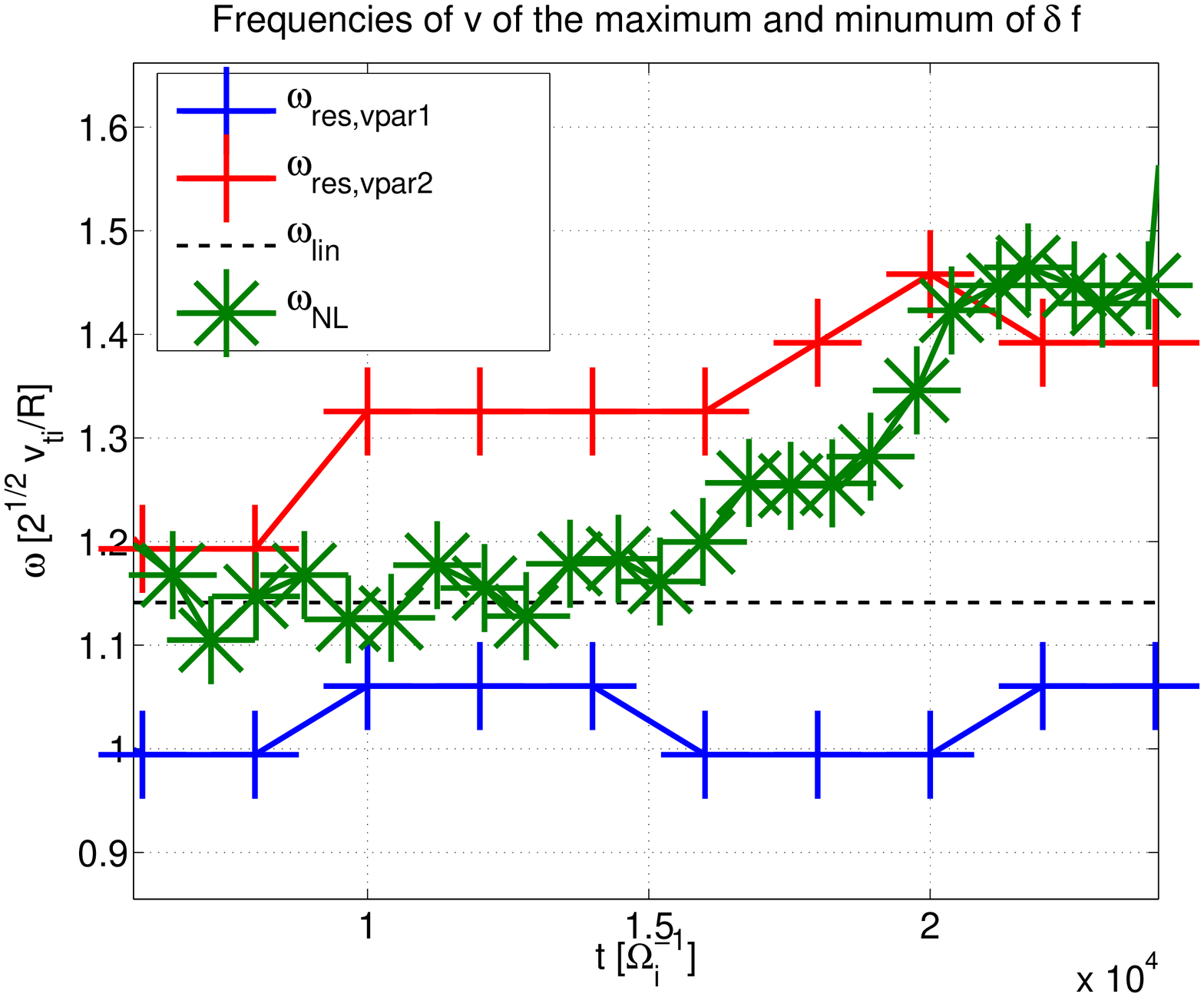}
\vskip -1em
\caption{Perturbed distr. funct. (left), for a case with $n_{EP}/n_i=0.176$.
The times are expressed here in units of $\Omega_i^{-1}$. On the right,
frequencies corresponding to the positive (red) and negative (blue) peaks of the EP perturbed distribution function. The measured EGAM freq. is also shown in green.\label{fig:nH17-chirp}}
\end{center}
\end{figure}

The EP distribution function of the case with $n_{EP}/n_i=0.176$ is shown in Fig.~\ref{fig:nH17-logabsE}-b, for different times up to the saturation. Firstly, we note that the range of velocities affected by the nonlinear modification is broader than that for the weaker drive. In fact, the redistribution of the EPs is observed to occur in a range of velocities between $2 \, v_{ti}$ and $5 \, v_{ti}$. Secondly, we note that the resonance frequency, which is calculated in this case from the linear frequency as $v_{\|res} = qR \, \omega_{EGAM} = 3.2 \, v_{ti}$, does not perfectly describe the velocity of the change of sign of the perturbed distribution function at all times. In fact, the resonance velocity is observed to grow in time, from $3.2 \, v_{ti}$ to $3.5 \, v_{ti}$.


The evolution of the resonance velocity in time is in relation with the EGAM nonlinear frequency modification, i.e. the EGAM chirping. The perturbed EP distribution function can be plotted explicitely (see Fig.~\ref{fig:nH17-chirp}-a). The positive peak (clump) and the negative peak (hole) can be seen to form and evolve in time, becoming bigger and centered at higher and higher distances from the linear resonance velocity.
The location in velocity space of the peaks can be measured and translated into resonance frequencies as $\omega_{1,2}=v_{\|1,2}/qR$ (see Fig.~\ref{fig:nH17-chirp}-b). When compared with the measured EGAM frequency, we note that the nonlinear EGAM frequency modification at the time of the saturation is described with a good approximation by the resonance frequency of the negative peak (hole).
This relation offers the possibility to predict the nonlinear frequency by approximating it with the frequency obtained by the velocity of the negative peak of the EP perturbed distribution function (see also Ref.~\cite{WangPRL13}).

In the next section, Sec.~\ref{sec:correspondence-BPS-EGAM}, we introduce the beam-plasma system (BPS) and the map linking the BPS with the EGAM. This map shows how we can predict the EGAM EP redistribution in velocity space.

\section{Dynamics of the energetic particles: analogy with the beam-plasma system}\label{sec:correspondence-BPS-EGAM}

In the EGAM system, the equilibrium magnetic field is not uniform but has a toroidal shape. In general, particles moving in a toroidal magnetic field can perform passing orbits, i.e. follow the magnetic field on both low-field side and high-field side, or perform banana orbits in the space restricted to the low-field side of the tokamak.
By construction, in the EGAM problem considered here, the energetic ions are initialized with a bump-on-tail distribution function (beam distribution), with a relatively high parallel mean velocity along the equilibrium magnetic field, plus a smaller isotropic thermal distribution around the mean velocity.
Due to the relatively high parallel velocity, the time derivative of their toroidal angle never vanishes (and therefore banana orbits of the EP in the low-field side are not considered in the present treatment). During their motion which is, to the leading order, directed along the equilibrium magnetic field, they perform small drifts towards higher values of the minor radius, and then towards lower values of the minor radius, known as the curvature and grad-B drifts. These drifts have zero time average: this defines orbits with an average radial position plus a radial orbit width. 

The radial electric field of the EGAM can exchange energy with the energetic particles, due to their radial component of the trajectories, and in particular of the curvature drift~\cite{Qiu11}: $v_{dc}=(v_\|^2/\Omega_{EP}) \, {\bf B}\times {\bf \nabla B}/B^2$.
Due to the structure of the perturbed distribution function, the effective parallel wavenumber of the EGAM is $k_\|=1/qR$, the phase-angle is $\Theta = \theta - \omega_{EGAM}t$ and its normalized time derivative is $\dot\Theta/\omega_{EGAM} = (v_\|-v_{\|r})/qR \, \omega_{EGAM}$. In terms of the phase angle, the energetic particles experience a periodic electric field, and their harmonic motion can be expressed as~\cite{Qiu11}:
\begin{equation}
\frac{\ddot{\Theta}}{\omega_{EGAM}^2} = - \frac{\omega_b^2}{\omega_{EGAM}^2} \sin\Theta\;,
\end{equation}
where $\omega_b$ is the bounce frequency of the energetic particles in the potential created by the wave. Using these considerations, the analogy with respect to the 1D beam-plasma system (BPS) turns out to be evident. In fact, the BPS is described as given by a Langmuir wave, excited by a beam of energetic electrons along a given direction $x$.
The Langmuir wave has a perturbed electric field directed along $x$,  oscillating at the plasma frequency $\omega_p = \sqrt{n_e e^2/m_e \epsilon_0}$. In general, the Langmuir wave can be decomposed in Fourier in terms of the wavenumbers $k_\ell=\ell(2\pi/L)$, where $L$ is the periodicity length of the 1D space domain of the BPS, and $\ell$ is a positive integer (whereas the EGAM has only one possible wavenumber set by the equilibrium, as mentioned above). Considering a single monochromatic wave with a chosen value of $\ell$, we can focus on that, and we denote the wavenumber as $k$. The phase-angle $\Theta$ experienced by the electrons in the field of the Langmuir wave is thus $\Theta=k x - \omega_p t$ and the normalized variation in time of the phase angle is $\dot\Theta/\omega_p = k (v-v_r)/\omega_p = k v/\omega_p - 1$, where the resonant velocity is defined as $v_r=\omega_p/k$. As a choice of nomenclature, we refer here to the velocities of the EGAM as $v_\|$, and to the velocities of the BPS as $v$. Similarly, we refer to the wavenumbers of the EGAM as $k_\|=1/qR$, and to the wavenumbers of the  BPS as $k$.

In the following, we describe the detailed mapping procedure which links the EGAM framework with the BPS. As already mentioned, the dynamics of the EGAM model can be reduced in the parallel velocity direction and we start from the generic resonance condition written using two suitable normalization constants $\nu_{1,2}$,
\ie
\begin{align}
\frac{v_{\|}-v_{\|0}}{\nu_1}=\frac{v-v_r}{\nu_2}\;,
\end{align}
where we recall that the transit resonance velocity reads $v_{\|res}=qR\omega_{EGAM}(n_{EP})$. Using the introduced above standard normalization $\nu_1=v_{ti}$ and, for the calculation of this Section, $\bar{v}_{\|}=v_{\|}/v_{ti}$, in the following, we denote with $\bar{v}_{\|min}\leqslant\bar{v}_{\|}\leqslant\bar{v}_{\|max}$ the domain of the positive bump of energetic particles.
Imposing the boundary  $\bar{v}_{\|min}=0\;\mapsto\;v_{min}=0$, in order to map one single bump of energetic particles with $\bar{v}_{\|}\geqslant0$, we get $\nu_2=v_r/\bar{v}_{\|r}$ and the map finally writes
\begin{align}
v=\frac{v_r}{\bar{v}_{\|r}}\,\bar{v}_{\|}\;.	
\end{align}
Let us now introduce the following normalization: $v=\omp(2\pi/L)^{-1}\,u$. In order to fix the dimensionless resonant wavenumber $\ell_r$, we use the condition $k_1v_{max}=\omp$, with $v_{max}=v_r\bar{v}_{\|max}/\bar{v}_{\|r}$, which characterize the spectral features (wavenumbers and periodicity length).
This yields to $\ell_r=\ell_1\bar{v}_{\|max}/\bar{v}_{\|r}$, and $\ell_r$ is determined arbitrarily fixing $\ell_1$ since $\bar{v}_{\|max}$ and $\bar{v}_{\|r}$ are given quantities from the EGAM system. We stress how the resonance condition can be rewritten as $\ell_r u_r=1$.

The map between the velocities of the two systems is now closed. The bump (positive part) of the EP is described by the shifted Maxwellian distribution function $F_{EP}(v_{\|})$ in velocity space~\cite{BiancalaniJPP17}.
For modelling the EP distribution function of the EGAM in the BPS,
let us now discretize the positive bump of $F_{EP}(v_{\|})$ in $n$ delta-like beams, equispaced in velocity space and located in $\bar{v}_{\|j}$ (with $j=1,\,...,\,n$), and assign the numbers of particles $N_j$ for each beam distributed according to $F_{EP}$. The initial conditions on the distribution for BPS simulations are now given by $N_j$ particles located at
\begin{align}\label{normmap}
u_j=\bar{v}_{\|j}/(\ell_r\bar{v}_{\|r})\;.
\end{align}
For the sake of completeness, we mention that, for the simulation of the BPS, we have set $n=600$, $\ell_1=400$ and we have used $N=10^6$ total particles. The complete derivation of the BPS dynamical equation used here, is described in \cite{Carlevaro15,Carlevaro16} (and refs. therein), and it can be specified for one single resonance as:
\begin{equation}\label{mainsys1}
\bar{x}_i'=u_i \;,\qquad
u_i'=i\,\ell_r\;\bar{\phi}_r\;e^{i\ell_r\bar{x}_{i}}+c.c.\;,\qquad
\bar{\phi}_r'=-i\bar{\phi}_r+\frac{i\eta}{2\ell_r^2 N}\sum_{i=1}^{N} e^{-i\ell_r\bar{x}_{i}}\;,
\end{equation}
where the particle position along the $x$ direction is labeled by $x_i$, with $i=1,\,...\,N$ ($N$ being the total particle number) normalized as $\bar{x}_i=x_i(2\pi/L)$.
The Langmuir electrostatic scalar potential $\phi(x,t)$ is expressed in terms of the Fourier component $\phi_r(k,t)$ and we have used: $\eta=n_B/n_p$ (for the plasma density $n_p$ assumed much greater than the beam one $n_B$), $\tau=t\omp$ (the prime indicates derivative with respect to this variable),  $\tilde{\phi}_r=(2\pi/L)^2 e\phi_r/m\omp^2$, $\bar{\phi}_r=\tilde\phi_r e^{-i\tau}$.
\erefs{mainsys1} are solved using a Runge-Kutta (fourth order) algorithm. For the considered time scales and for an integration step $h=0.1$, both the total energy and momentum (for the explicit
expressions, see \cite{Carlevaro15}) are conserved with relative fluctuations of about $1.4\times10^{-5}$.

The BPS is closed once the density of the beam (drive) is fixed. In order to quantitatively compare the non-linear features of two systems, we now fix the bounce (trapping) frequency $\omega_B$ normalized to the mode frequency equal for the two schemes. For the BPS, the bounce frequency results proportional to the linear growth rate of the mode $\gamma_{L,\bb}$. The same occurs for the EGAM system, but with a proportionality factor depending on the EP density~\cite{BiancalaniJPP17}. In particular, we get:
\begin{align}
\frac{\omega_{B,\bb}}{\omp}=\alpha\,\frac{\gamma_{L,\bb}}{\omp}\;,\qquad
\frac{\omega_{B,\ee}}{\omega_{L,EGAM}}=\beta(\bar{n}_{EP})\,\frac{\gamma_{L,\ee}}{\omega_{L,EGAM}}\;.
\end{align}
where $\alpha\simeq3.3$ (see well-known literature results \cite{ZCrmp,wu95} and also \cite{Carlevaro18}) while for the EGAM we have $\beta = \beta_0 \sqrt{\omega_{L,EGAM}/\omega_{GAM}}$, with $\beta_0=2.66$ in this regime, and $\omega_{L,EGAM}$ depending on the EP density~\cite{BiancalaniJPP17}. For the  four selected EGAM simulations, we get $\beta(0.07,\,0.10,\,0.176,\,0.30)\simeq[2.21,\,2.17,\,2.07,\,1.98]$. Using standard normalization for frequencies, \ie $\bar{\gamma}_{L,\bb}=\gamma_{L,\bb}/\omp$ and $\bar{\gamma}_{L,\ee}=\gamma_{L,\ee}/\omega_{GAM}$, $\bar{\omega}_{L,EGAM}=\omega_{L,EGAM}/\omega_{GAM}$ and equaling the bounce frequencies, we finally get
\begin{align}\label{gl}
\bar{\gamma}_{L,\bb}=\frac{\beta}{\alpha}\,\frac{\bar{\gamma}_{L,\ee}}{\bar{\omega}_{L,EGAM}}\;.
\end{align}

This condition preserves the linear and nonlinear features of the two systems and it is used in the evaluation for the drive of BPS simulations from the linear dispersion relation which formally reads as
\begin{align}\label{jjdbnfhjkb}
\epsilon=1-\frac{\omp^2}{\omega^2}=\frac{\eta \omp^2}{k^2}
\int_{-\infty}^{+\infty}dv\frac{k\,\p_v \hat{F}_B(v)}{k v-\omega}\;.
\end{align}
where $\hat{F}_B(v)$ is the initial beam distribution function. Here, the dielectric function $\epsilon$ can be expanded near $\omega\simeq\omp$ to deal with Langmuir modes as in \erefs{mainsys1}, \ie $\epsilon\simeq2(\bar{\omega}-1)$ (where $\bar{\omega}=\omega/\omp$).
Let us now use the expansion $\bar{\omega}=\bar{\omega}_0+i\bar{\gamma}_{L,\bb}$, where $\bar{\omega}_0$ is the real part  of the normalized Langmuir frequency $\bar\omega$. Using the linear character of the mapping  which yields to the normalization $\hat{F}_B(v)=\kappa F_{EP}(\bar{v}_{\|})$ (with $\kappa=const.$), \eref{jjdbnfhjkb} can be written in terms of the EGAM system variables as 
\begin{align}
2(\bar{\omega}_0+i\bar{\gamma}_{L,\bb}-1)-\frac{\eta\bar{v}_{\|r}}{M}
\int_{-\infty}^{+\infty}\!\!\!\!\!\!\!d\bar{v}_{\|}\frac{\p_{\bar{v}_{\|}}
F_{EP}}{\bar{v}_{\|}/\bar{v}_{\|r}-\bar{\omega}_0-i\bar{\gamma}_{L,\bb}}=0\;,
\end{align}
where $M=\int_{-\infty}^{+\infty} d\bar{v}_{\|}F_{EP}$. This equation is numerically integrated assuming \eref{gl}, which guarantees the requested features described above, and provides the drive parameter $\eta$ closing the map procedure.

\section{Nonlinear EP redistribution of the EGAM, and comparison with the beam-plasma model}
\label{sec:EP-redistribution}

\begin{figure}[b!]
\begin{center}
\includegraphics[width=0.47\textwidth]{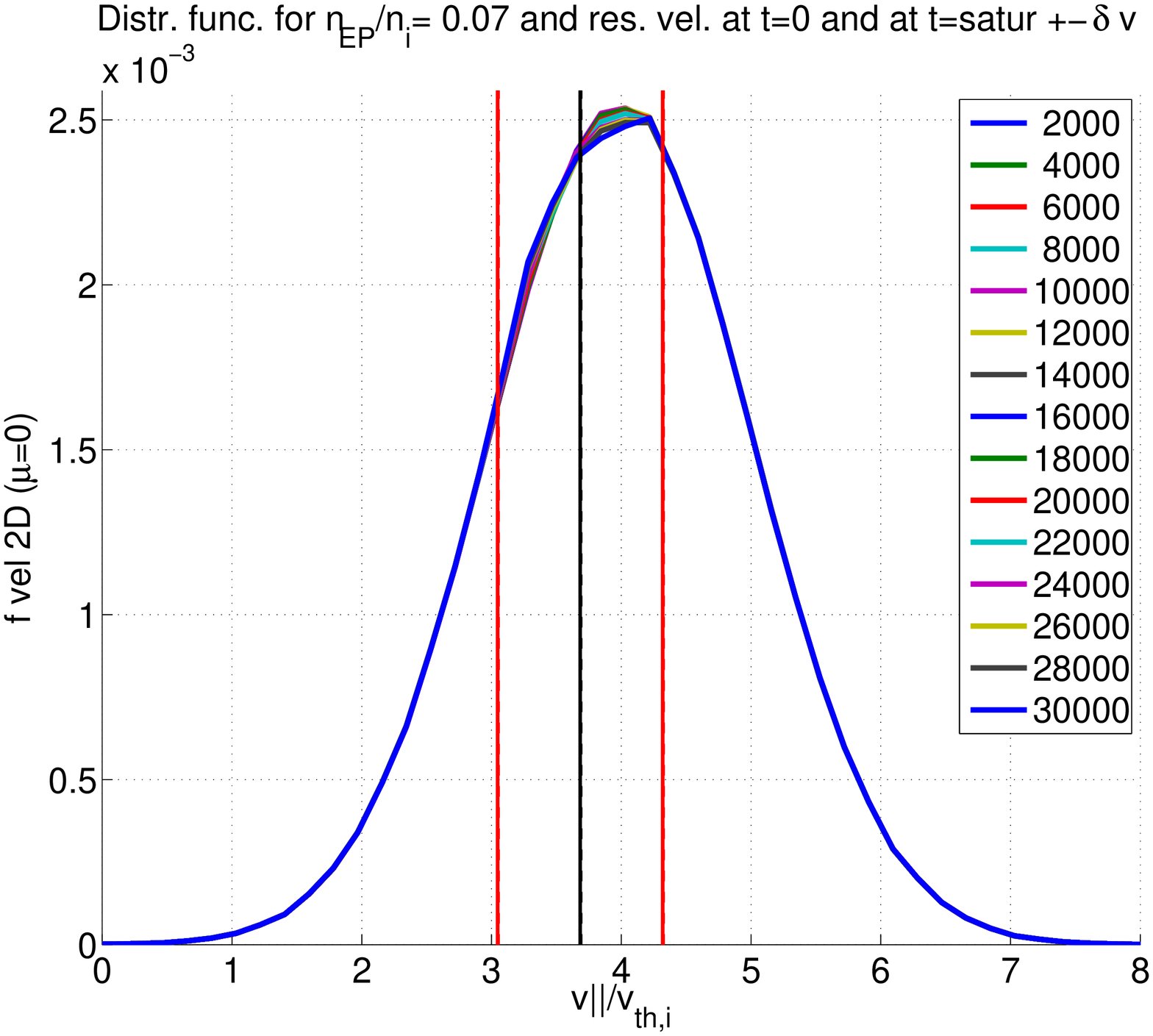}
\includegraphics[width=0.47\textwidth]{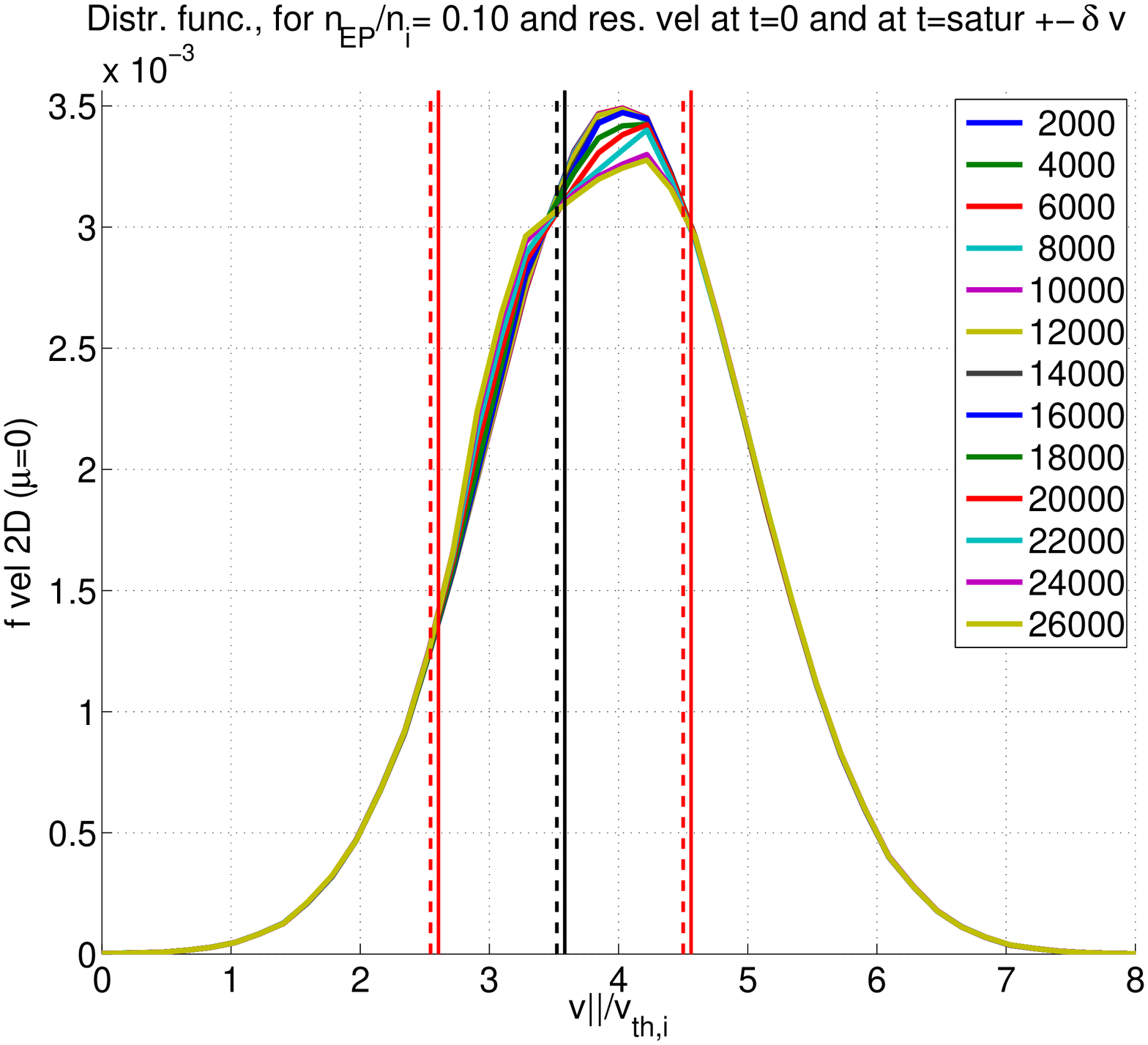}
\caption{Energetic particle distribution function averaged in space, and measured at $\mu\simeq 0$, vs parallel velocity, for $n_{EP}/n_i=0.07$ (left-hand panel) and $n_{EP}/n_i=0.10$ (right-hand panel). The vertical dashed and continuous lines are the resonance velocity (center), with the borders of the nonlinear velocity predicted by $v_{\|res}\pm\Delta v_{\|NL}$.\label{fig:distr-funct-n07-n10}}
\end{center}
\end{figure}

In this section, we compare the results of the EP redistribution due to the EGAM, with the redistribution due to the beam-plasma system. In particular, we are aimed at predicting, from BPS informations, the nonlinear parallel velocity spread in the positive bump of the distribution function.

In the BPS, the single mode dynamics proceeds in an initial exponential mode growth followed by non-linear saturation. Here the particles get trapped and begin to bounce back and forth in the potential well generating clumps. A measure of the clumps width $\Delta{u}^{c}_{NL}$ for a generic initial half-Gaussian velocity distribution, which can be directly extrapolated to the analysis of the present work, has been evaluated in \cite{Carlevaro18} for several cases outlining the following scaling rule as function of the linear drive:
\begin{equation}
\Delta u^{c}_{NL}/u_r = (6.64\pm0.12)\;\bar{\gamma}_L.
\end{equation}
In order to include the dynamic role also of  passing but nearly resonant particles \cite{EEbook}, i.e. the region involved in the effective wave particle power exchange, in the following analysis we consider as the proper nonlinear particle velocity spread the scaled quantity $\Delta u_{NL}\simeq \chi \Delta u^{c}_{NL}$ with $\chi\simeq1.28$.
This estimate is derived \cite{Carlevaro18} characterizing the active overlap of different non-linear fluctuations \cite{ED81,LL10} and corresponds to the finite distortion of the distribution function, including effects at the edges of the plateau (defined as the flattened region of the distribution function, mainly coinciding with the clump size). We finally obtain the desired formula for the prediction of the nonlinear velocity spread due to the EGAM:
\begin{equation}\label{eq:NL-v-spread-1}
\frac{\Delta v_{\|NL}}{v_{\|L,res}} = 8.5 \; \bar{\gamma}_L^{BPS} 
\end{equation}
and by substituting the value of $\bar\gamma_L^{BPS}$ we obtain:\begin{equation}\label{eq:NL-v-spread-2}
\frac{\Delta v_{\|NL}}{v_{\|L,res}}= 2.57 \, \frac{\beta_0}{\sqrt{\omega_{GAM}}} \frac{\gamma_L^{EGAM}}{\sqrt{\omega_{L,EGAM}}} 
\end{equation}
Eq.~\ref{eq:NL-v-spread-2} has been derived for the EGAM system, using the scaling derived in \eref{gl} and the normalized mapping of Eq.~\reff{normmap}.
For the regime of interest in this paper, we have $\beta_0=2.66$~\cite{BiancalaniJPP17}, and $\omega_{GAM}= 1.8 \, \omega_s$, therefore we obtain:
\begin{equation}\label{eq:NL-v-spread-3}
\frac{\Delta v_{\|NL}}{v_{\|L,res}} = \frac{5.1}{\sqrt{\omega_s}} \frac{\gamma_L^{EGAM}}{\sqrt{\omega_{L,EGAM}}}
\end{equation}

\begin{figure}[t!]
\begin{center}
\includegraphics[width=0.47\textwidth]{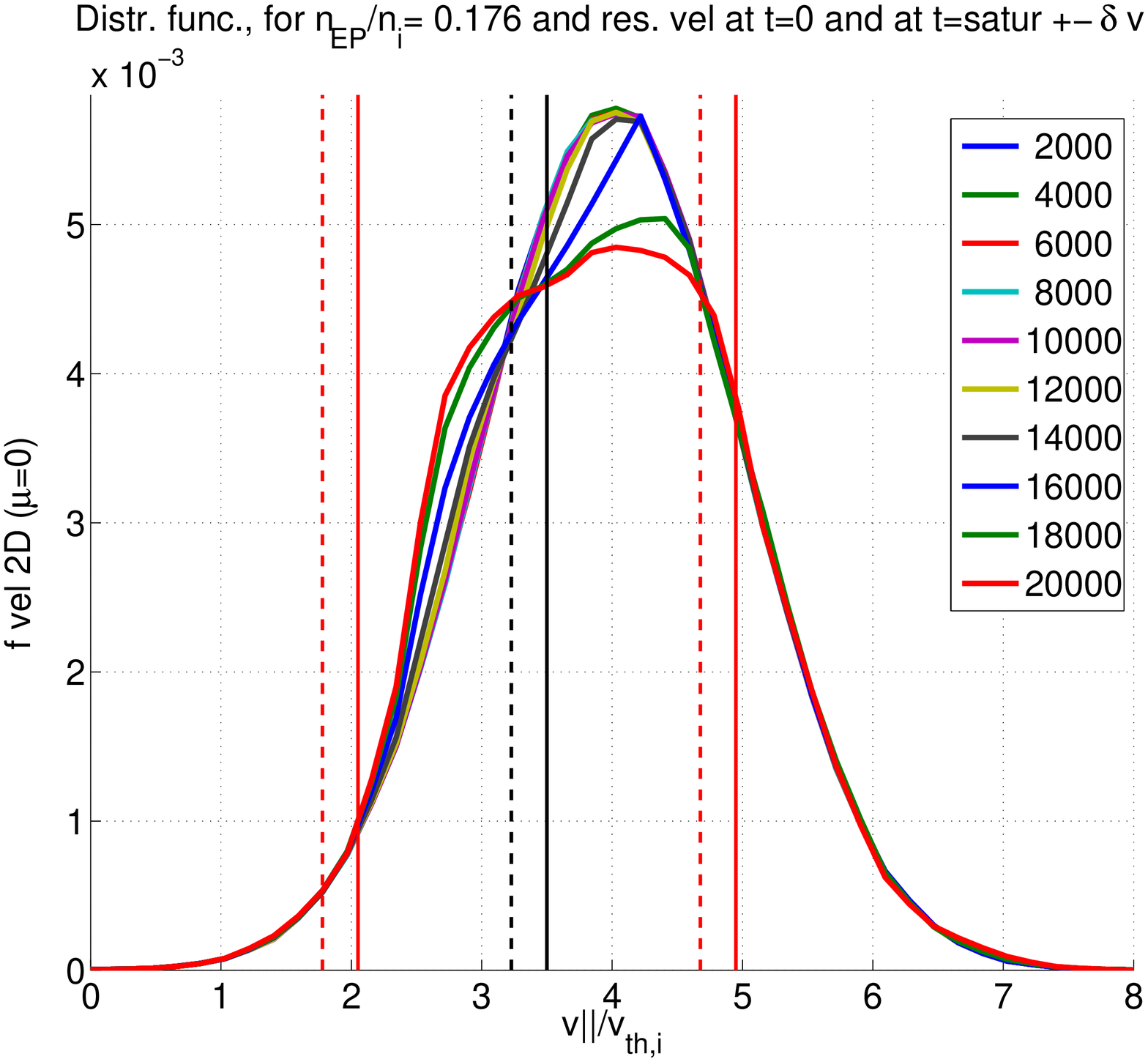}
\includegraphics[width=0.47\textwidth]{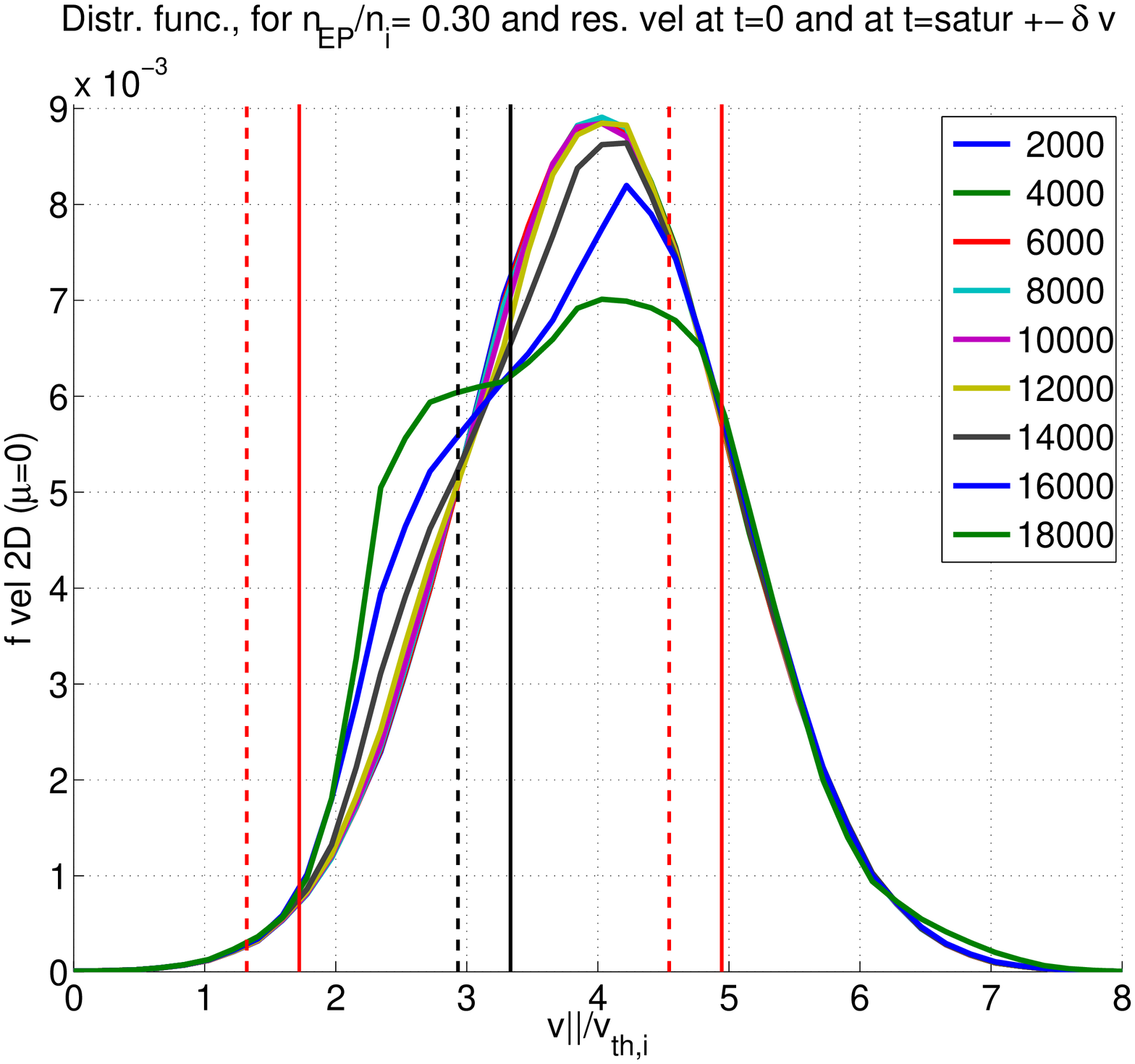}
\caption{Same as \figref{fig:distr-funct-n07-n10} but for $n_{EP}/n_i=0.176$ (left panel) and $n_{EP}/n_i=0.30$ (right panel).\label{fig:distr-funct-n17-n30}}
\end{center}
\end{figure}

Let us now analyze the predictivity of Eq.~\ref{eq:NL-v-spread-2}, simplified as in Eq.~\ref{eq:NL-v-spread-3}  for the regime of interest. Four different simulations are considered, with different values of energetic particle concentration: $n_{EP}/n_i\in[0.07,\,0.10,\,0.176,\,0.30]$. The corresponding linear frequencies and growth rates are $\omega_{L,EGAM}= [1.30,\, 1.24,\, 1.14,\, 1.04] \omega_s$ and $\gamma_{L,EGAM}= [0.04,\, 0.06,\, 0.09,\, 0.11] \omega_s$.
In all simulations, the distribution function is averaged in space, and measured at $\mu\simeq 0$. Snapshots of the distribution function at different times are selected, going from the linear phase to the instant when the first peak of the electric field is reached, i.e. at saturation.
In \figref{fig:distr-funct-n07-n10} and \figref{fig:distr-funct-n17-n30}, the distribution functions measured with ORB5 are shown to be modified by the nonlinear interaction with the EGAM, with a certain width in the velocity space. The position of the linear resonance velocity and the nonlinearly modified resonance velocity are shown for each simulation respectively as a dashed vertical black line, and a continuous vertical black line.
We can now compare with the predicted values of Eq.~\ref{eq:NL-v-spread-3}: $\Delta v_{\|NL}/v_{\|L,res} = [0.17,\, 0.28,\,  0.45,\, 0.55]$. The corresponding predicted range is delimited by vertical red lines in \figref{fig:distr-funct-n07-n10} and \figref{fig:distr-funct-n17-n30}. The dashed lines correspond to the range calculated with respect to the linear resonance velocity, and the continuous lines with respect to the nonlinearly modified resonance velocity.
Note that the predicted width of the velocity domain of EP redistribution, centered at the nonlinearly modified resonance velocity, fits very well with the results of ORB5.
Note also that, for the two cases with lowest drive, the nonlinear modification of the resonance velocity is negligible, and therefore the predicted width of the velocity domain of EP redistribution fits very well the results of ORB5, even when centered at the linear resonance velocity.

\begin{figure}[t!]\centering
\includegraphics[width=0.45\textwidth,clip]{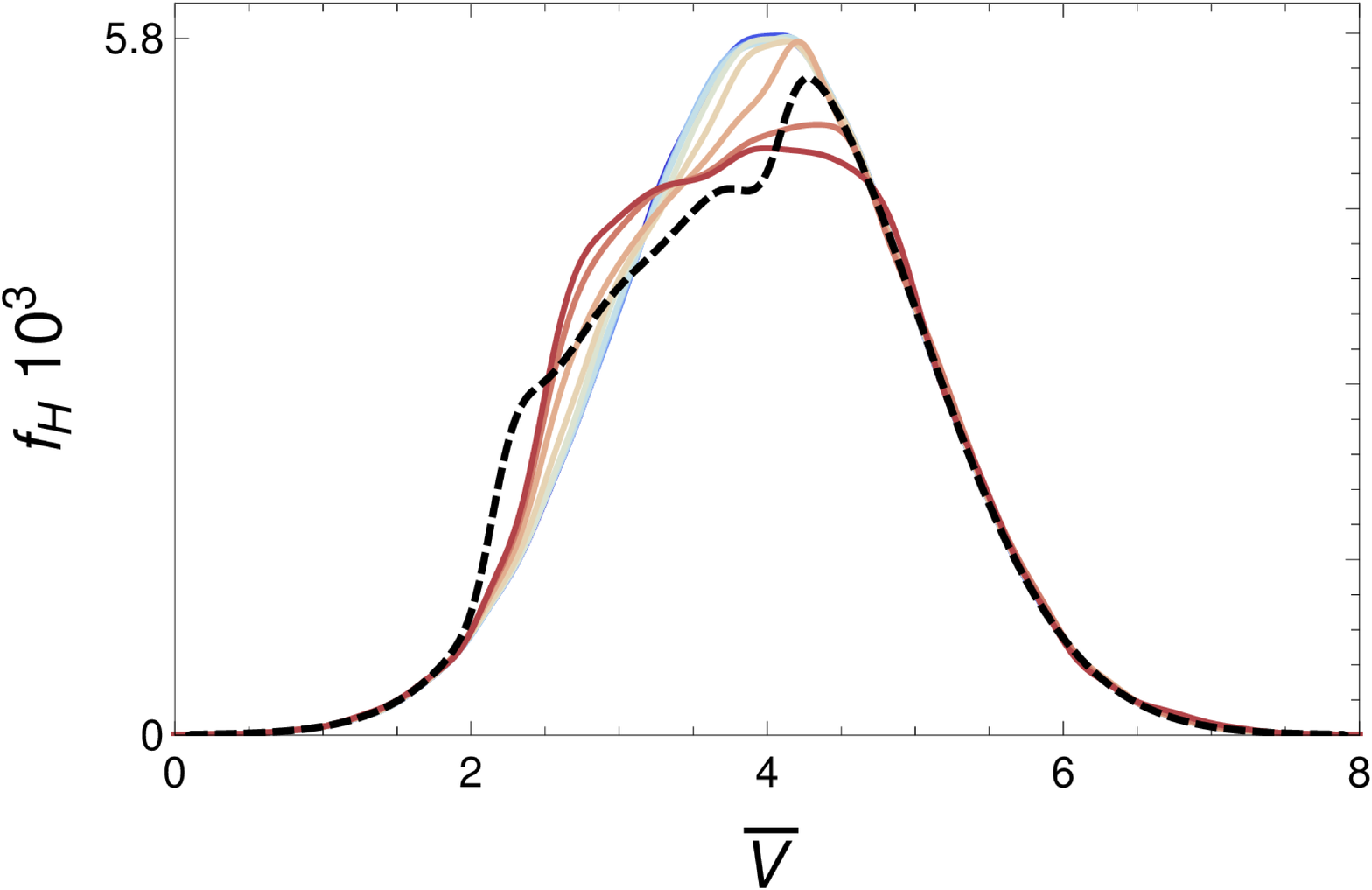}
\includegraphics[width=0.45\textwidth,clip]{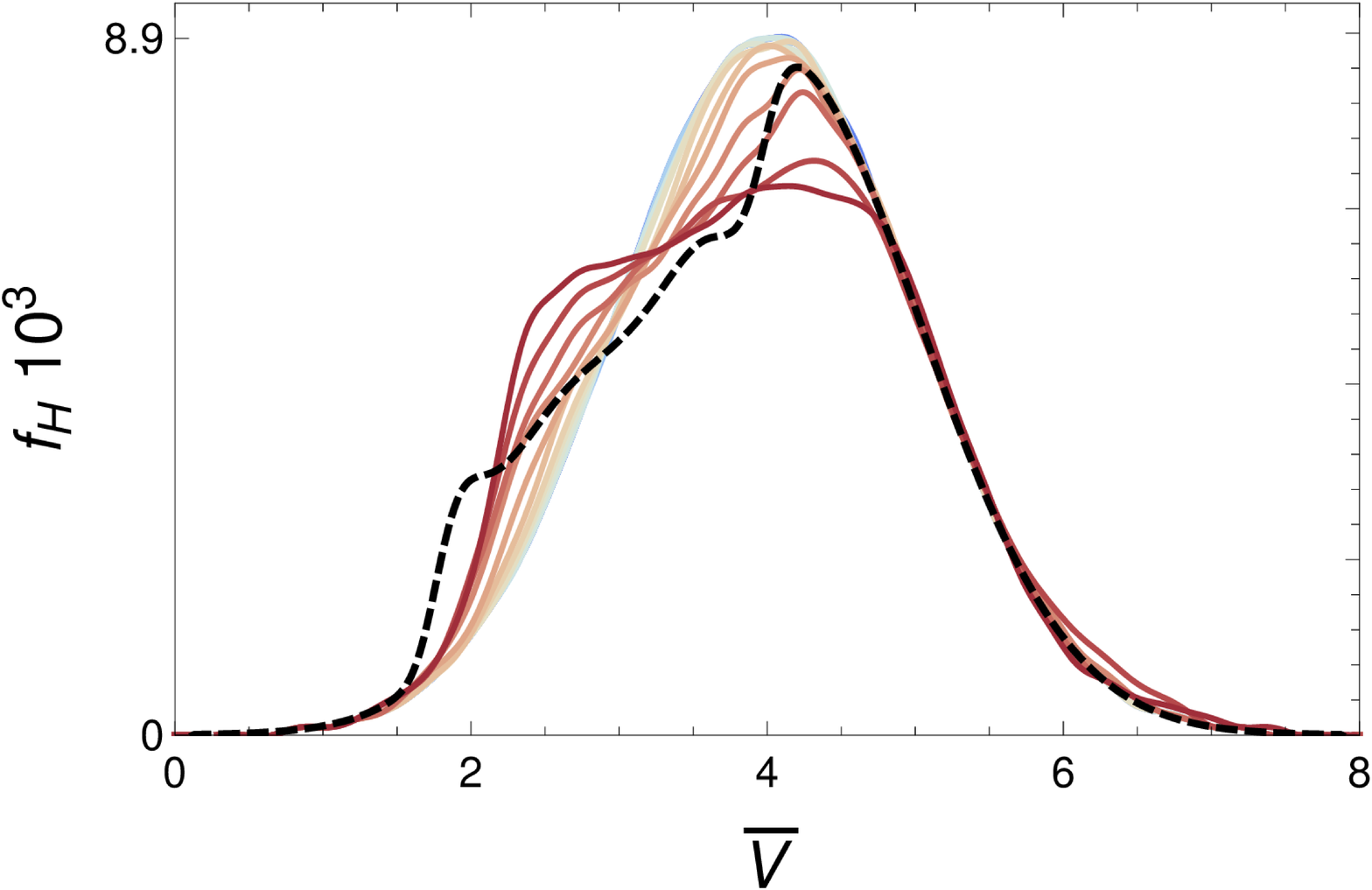}
\caption{Plot of the BPS distribution function at saturation (dashed-black) mapped back in the $v_{\|}$ space, over the evolution of the EP profile for $n_{EP}/n_i=0.176$ (left-hand panel) and $n_{EP}/n_i=0.3$ (right-hand panel).\label{fig2}}
\end{figure}

In summary, two distinct regimes can be identified. When the instabilities are weakly driven (\figref{fig:distr-funct-n07-n10}), a very good match between the estimated deviation $v_{\|res}\pm\Delta v_{\|NL}$ and the nonlinear EP redistribution is observed, where $v_{\|res}$ is the linear resonance velocity. Otherwise, in the strongly driven regime (\figref{fig:distr-funct-n17-n30}) the importance of the frequency chirping comes out. In particular, nonetheless the value $v_{\|res}\pm\Delta v_{\|NL}$ remains very predictive, as long as the nonlinearly modified resonance velocity is chosen as center of the EP redistribution due the EGAM. 

The nonlinear frequency shift which is characteristic of the EGAM, can not be intrinsically implemented in the BPS model. In fact, differences remain in the physics of the EGAM and the BPS.
This clearly emerge in \figref{fig2}, where the distribution function of the BPS at saturation (dashed-black line), mapped back in the $v_{\|}$ space, is overplotted on the evolution of the EP profile for $n_{EP}/n_i=0.176$ and $n_{EP}/n_i=0.3$.
In particular, it is evident how the discrepancy due the fixed character (at $\sim\omp$) of the Langmuir resonance gives rise to a very different morphology of the distribution function, although well predicting the effective nonlinear velocity spread.
Also the inclusion of additional modes with artificial \emph{ad hoc} damping rates results in a drastically non-comparable non-linear dynamics, underlining the intrinsic differences of the physical systems.

\section{Conclusions and discussion}\label{sec:conclusions}

Geodesic acoustic modes (GAMs), i.e. finite frequency zonal (i.e. axisymmetric) flows with mainly radial electric field polarization, are known to be important in tokamaks due to their interaction with turbulence. GAMs can also be excited by energetic particles (EPs), taking the name of EGAMs. In the view of understanding and predicting the EP redistribution in tokamaks, the linear and nonlinear dynamics of EGAMs should be properly theoretically understood.

In this paper, we have investigated the nonlineary dynamics of EGAMs with particular interest in the EP redistribution in velocity space. As a tool of investigation of the nonlinear inverse Landau damping, which is responsible of the EGAM saturation and EP redistribution, a comparison of the EGAM dynamics with the beam-plasma system (BPS) is done. Although the BPS describes the interaction of EP with a different mode, i.e. the Langmuir wave in a 1D geometry, nevertheless, analogies with the EGAM had been suggested in previous papers. These analogies are investigated here and used to build a mapping of the two systems. The mapping is used to understand and predict the EP redistribution of the EGAM.

The EGAM is investigated here with collisionless electrostatic simulations with the gyrokinetic (GK) particle-in-cell code ORB5. Only the wave-particle nonlinearity is retained in the simulations presented here, meaning that the markers for thermal ions follow unperturbed trajectories in phase space. The GK model allows to study the EGAM dynamics retaining the crucial physics of the resonances with thermal and fast ions. The BPS is investigated with the a 1D code treating the thermal plasma as a cold dielectric medium and describing the dynamics of the fast particles (fast electrons in the case of the BPS) as an N-body problem solved with an Hamiltonian formulation.

The GK simulations of the EGAM show that, after a first linear growth, the EGAM enters a first nonlinear phase where the EP distribution function suffers a modification due to the EGAM field. A saturation of the EGAM field occurs, and then a deep nonlinear phase is entered, where nonlinear oscillations of the fields are observed. In this paper, we are interested in the first nonlinear phase only, up to the first saturation. The EP population is observed to redistribute, with EP going from higher to lower values of the parallel component of the velocity, as the EGAM grows in amplitude. The resonant velocity is also measured with ORB5 in the different cases considered.

The mapping of the EGAM and the BPS is then described, and the comparison of the EP redistribution in velocity space is shown. In particular, the main result is the prediction of the width of the velocity space $\Delta u_{NL}$ which is affected by the EGAM, around the resonance. The implication of this result is evident, as reduced models are needed for predicting the nonlinear dynamics of instabilities in tokamaks, instead of using numerically expansive GK simulations. In fact, BPS simulations are numerically much cheaper, and therefore the mapping described here offers a tool for predicting the EP redistribution in regimes where EGAMs experience wave-particle nonlinear saturation.

We have also found a transition among two regimes: for weakly driven EGAMs, the resonant velocity does not evolve in time during the first nonlinear phase, exactly like in the BPS problem; on the other hand, above a certain threshold in drive, the resonant velocity slightly increases in time, and therefore a difference of the EGAM and BPS is found. The increase of the resonant velocity in the high-drive regime is consistent with the nonlinear frequency modification (i.e. the frequency chirping) which is present for the EGAM, and is absent for the BPS considered here. The difference can be cured by calculating the nonlinear width $\Delta u_{NL}$ around the new value of the resonant velocity. This proves that the one-to-one correspondence of the EP redistribution \emph{around} the resonance is completely described by the nonlinear inverse Landau damping which is included in the model of the BPS. Therefore, we can state that the EP are redistributed around the resonance by the EGAM for a purely 1D problem which is the same as the BPS, namely the nonlinear inverse Landau damping. Having tried to include many modes in the BPS, we have observed that this can modify the EP redistribution, but leads to a overestimation of the EP redistribution due to a lack of damping in the different modes. 

In the view of the continuation of this work, several next steps can be done in the direction of getting closer to more and more realistic scenarios. For example, the inclusion of wave-wave coupling of the EGAM with itself is under investigation. Kinetic electron effects should also be considered, as the electron resonances might affect the Landau damping. The inclusion of turbulence is also in progress, as a mean of modifying the EGAM saturation level and EP redistribution.

\section*{Acknowledgements}

Useful discussions with F. Zonca, P. Lauber, D. Zarzoso, I. Novikau, A. Di Siena, \"O. G\"urcan and P. Morel are acknowledged. Useful discussions with L. Villard, and the whole ORB5 team, are also acknowledged.
Part of this work has been carried out within the framework of the EUROfusion Consortium and has received funding from the Euratom research and training programme 2014-2018 under grant agreement No 633053, within the framework of the {\emph{Nonlinear energetic particle dynamics}} (NLED) European Enabling Research Project, WP 15-ER-01/ENEA-03 and {\emph{Nonlinear interaction of Alfv\'enic and Turbulent fluctuations in burning plasmas}} (NAT) European Enabling Research Project, CfP-AWP17-ENR-MPG-01. The views and opinions expressed herein do not necessarily reflect those of the European Commission.
Simulations were performed on the Marconi supercomputer within the framework of the OrbZONE and ORBFAST projects. Part of this work was done while one of the authors (A. Biancalani) was visiting ENEA-Frascati, whose team is acknowledged for the hospitality. This paper, together with the companion paper on EGAM saturation and comparison with the beam-plasma-system (i.e. Ref.~\cite{BiancalaniJPP17}), are dedicated to Prof. Francesco Pegoraro.

\end{document}